\documentclass[twocolumn,numbering,showpacs]{revtex4-1}

\usepackage{graphicx,color}
\usepackage{latexsym}
\usepackage{amsmath,amssymb}        
\usepackage[draft=false]{hyperref}
\usepackage{mathrsfs}
\usepackage{comment}
\usepackage{soul}
\usepackage{mathrsfs}
\definecolor{purple}{rgb}{0.58,0.0,0.83}
\definecolor{orange}{rgb}{1,0.5,0}
\DeclareSymbolFontAlphabet{\mathrsfs}{rsfs}
\DeclareMathAlphabet{\mathcal}{OMS}{cmsy}{m}{n}

\begin{document}

% -----> TITLE 

\title{Merger of galactic cores made of ultralight bosonic dark matter}

% -----> AUTHORS 

\author{F. S. Guzm\'an, J. A. Gonz\'alez, I. Alvarez-R\'ios}
\affiliation{Laboratorio de Inteligencia Artificial y Superc\'omputo, Instituto de F\'{\i}sica y Matem\'aticas, Universidad Michoacana de San Nicol\'as de Hidalgo. Edificio C-3, Cd. Universitaria, 58040 Morelia, Michoac\'an, M\'exico.}
% --->   DATE

\date{\today}

% -----> ABSTRACT

\begin{abstract}
We study binary mergers of ultralight bosonic dark matter cores by solving the Gross-Pitaevskii-Poisson system of equations. The analysis centers on the dynamics of the relaxation process and the behavior of the configuration resulting from the merger, including the Gravitational Cooling with its corresponding emission of mass and angular momentum. The oscillations of density and size of the final configuration are characterized, indicating that for the equal mass case the dependency of the amplitude and frequency of these oscillations on the impact parameter of the pre-merger configuration is linear. The amplitude of these oscillations changes by a factor of two or more indicating the final configuration does not approach a clear stationary state even though it oscillates around a virialized state. For the unequal mass case, global quantities also indicate the final configuration oscillates around a virialized state, although the density does not show a dominant oscillation mode. Also the evolution of the angular momentum prior and post merger is analyzed in all cases. 
\end{abstract}

% ----->   PACS
\pacs{}

% ----->   MAKETITLE 

\maketitle

% -------------------------------------
% ----->     SECTION    <-----
% -------------------------------------
\section{Introduction}
\label{sec:introduction}

One of the viable dark matter candidates currently under study is the ultralight spin-less boson \cite{Matos-Urena:2000,Ostriker:2016}, which is attractive because of some interesting properties consistent with observations. For instance when its mass is of order $m\sim 10^{-22}eV$ structures do not develop cusps due to the large de Broglie length \cite{Chen:2016,Schive:2015,Du:2016,Velmaat2018} whereas at large scale the behavior is consistent with that of CDM \cite{Schive:2014,Schive:2014hza}. At the same time, this model is also consistent with the small structure abundance of the mass power spectrum \cite{Matos-Urena:2000,Ostriker:2016,Marsh-Ferreira:2010,Schive:2015}.

Local scale dynamics on the other hand, should indicate  differences between CDM and ultralight bosoinc dark matter and impose constraints on the later. For instance, the relaxation process should be special, being the gravitational cooling process an option \cite{GuzmanUrena2003,GuzmanUrena2006} in which matter carries out kinetic energy leaving the structure under relaxation in a nearly virialized state, or other processes involving dynamical friction \cite{Mocz2017}, or damping   \cite{ChavanisDamping} could provide the relaxation mechanism. Also the collisions and interaction between structures can provide important restrictions to the model, for example
the density resulting from head-on core mergers \cite{AvilezGuzman2019} that may result in the destruction of luminous matter clusters during the process for certain particular scenarios \cite{GonzalezGuzman2016}.
Other restrictions, this time on the boson mass are found from the analysis of core oscillations that may or may not allow the formation of star clusters in galaxies \cite{Marsh2018}.

Locally, the dynamics of this dark matter model is ruled by the Gross-Pitaevskii-Poisson (GPP) system, that describes the evolution of a Bose-Einstein Condensate in the Gross-Pitaevskii mean field approximation, contained by the gravitational potential generated by itself. One point the various studies and approaches at local scale of the model have in common,  is that this type of dark matter clumps into structures with a universal profile, either into an equilibrium configuration of the GPP system for isolated systems \cite{GuzmanUrena2006,BernalGuzman2006b}, or composed of a core, sometimes called solitonic profile that matches the density profile of an equilibrium configuration \cite{Schive:2014}, and a surrounding cloud with a NFW profile obtained from simulations involving structure formation clustering \cite{Schive:2014,Schive:2014hza,Schwabe:2016,Mocz2017,Niemeyer2016,ChavanisCore}.

Among the common interactions between structures or cores, the merger of two of them is very important and is the subject of this paper. Configurations resulting from a merger with angular momentum, naturally inherit rotation from the original merging cores. Rotating structures within this dark matter model are interesting for various reasons. One of them is that rotation is an extra parameter for BEC dark matter halos that help fitting galactic rotation curves by keeping the boson mass unchanged \cite{BEC2014,BEC2015}, and will possibly help to reduce the dispersion of boson mass in rotation curve fitting of big catalogs \cite{Tula}. In a similar context, ellipsoidal analytic solutions to the GPP with rotation have been associated with possible vortex solution \cite{RindlerShapiro2012}. And more recently, new exact solutions of the GPP system with rotation are also being constructed with the aim of studying this dark matter model at local scale \cite{DavidsonSchwetz2016,Hertzberg2018}.

The study of core mergers in orbit or during structure formation, within the context of ultralight bosonic dark matter is not new. In fact also multiple soliton mergers have also been studied \cite{Schwabe:2016,ParedesMichinel2016,Mocz2017}. Specially in \cite{Schwabe:2016} the mergers have been analyzed in detail, from the initial conditions to the properties of the final configuration. Among the most interesting results it was found that the final mass of the merger does not depend on the initial momentum of the orbiting objects and only depends on mass ratio, total initial mass and total energy of the system. Also in \cite{Niemeyer2016} the density of cores resulting from mergers is compared with the solitonic profile in the context of structure formation simulations.

The analysis in our paper is very similar to that in \cite{Schwabe:2016}, however some new results arise.  Important differences are that we solve the GPP system without using the Madelung transformation, not for calculations nor for diagnostics of macroscopic quantities. We in fact confirm that the final mass of the merger does not depend on the initial angular momentum of the pre-merger configuration, however we find this result holds only for the equal mass case. We also find that the angular momentum of the final configuration depends on the initial conditions prior to merger, for both the equal and unequal mass cases. 

On the other hand, in \cite{Velmaat2018}  within the analysis of structure formation it is found that cores exhibit strong undamped oscillations. Our results are consistent with this evidence. From our analysis, we find that the configuration resulting from the merger of two cores exhibits a  dynamical behavior, characterized by oscillations with considerable amplitude that depend on the parameters of the binary system. The final structure does not relax,  however by fitting the density profile at different times we illustrate how the core radius and central density change in time.

The paper is written with the following structure. 
In Sec~\ref{sec:system} we describe the method used to simulate the mergers.
In Sec~\ref{sec:analysis} we analyze the equal and unequal mass scenarios.
In Sec~\ref{sec:conclusions} we draw some conclusions.

% -------------------------------------
% ----->     SECTION    <-----
% -------------------------------------
\section{Evolution of the system}
\label{sec:system}

Like in the analyses of structure formation and binary mergers mentioned before, we assume the dynamics of the ultralight bosonic dark matter is ruled by the GPP system of equations. Likewise we assume the free field regime, where the self-interaction among bosons is neglected, the so called fuzzy dark matter regime. Finally, we solve the equations using numerical methods and initial conditions described below.

\subsection{Numerical methods}

We solve the time dependent GPP system of equations which in code units is written as

\begin{eqnarray}
i\partial_t \Psi &=& -\frac{1}{2}\nabla^2 \Psi + V \Psi\nonumber \\
\nabla^2 V &=& |\Psi|^2,
\label{eq:gpp}
\end{eqnarray}

\noindent that describes the evolution of the fuzzy dark matter. Here, $\Psi$ represents the wave function of the system  and $|\Psi|^2$ is interpreted as the macroscopic density of the condensate and $V$ is the gravitational potential sourced by the condensate itself. We solve these equations for $\Psi$ in a cubic finite domain, with initial data for $\Psi$ consistent with the potential $V$. In this system, Poisson equation is a constraint that has to be solved on the fly as the bosonic gas density evolves.

We solve the Gross-Pitaevskii equation  numerically in 3D using the method of lines for the evolution across spatial slices separated by intervals of time $\Delta t$. 
The spatial domain $D=[x_{min},x_{max}]\times[y_{min},y_{max}]\times [z_{min},z_{max}]$ is described with a Cartesian and uniformly discretized grid defined by $x_{i,j,k} = x_{min}+i\Delta x$, $y_{i,j,k} = y_{min}+j\Delta y$ and $z_{i,j,k} = z_{min}+k\Delta z$, for $i=0,...,N_x$, $j=0,...,N_y$, $k=0,...,N_z$, with an isotropic resolution  $\Delta x = \Delta y = \Delta z = (x_{max}-x_{min})/N_x$. 

We discretize the equations with second order accurate finite difference stencils for spatial derivatives. For the sake of accuracy in the region of the merger, we use fixed mesh refinement based on the Berger-Oliger  algorithm \cite{BergerOliger}, with concentric refinement boxes. The resolution factor between successive refinement levels is one half. Considering that for the stability of the evolution, time and space resolution are limited by the condition $C=\Delta t / \Delta x^2 < 0.25 / \sqrt{3}$, we choose the value of $C$ to be that corresponding to the most refined level.

We solve Poisson equation for $V$ with a Multigrid algorithm with subcycles that use the Successive Over Relaxation method. This equation is solved at initial time and during the evolution. Due to its computational cost, the integration of Poisson equation  represents the major bottleneck of the code during the simulations.

Since we want to avoid reflections of matter from the boundary of the numerical domain, and because the Gravitational Cooling depends on the emission of matter that carries kinetic energy with it, we implement a sponge consisting of the addition of an imaginary potential such that $V\rightarrow V+V_{im}$, acting as a sink of particles following the recipe in \citep{GuzmanUrena2004}.  We make sure that the transition region of the sponge lies exclusively in the coarsest refinement level.

% ------------
\subsection{Initial conditions}

We assume the colliding objects are equilibrium configurations, which are spherical stationary solutions, constructed by assuming a harmonic time dependence of the wave function $\Psi=e^{-i\omega t}\psi(r)$, where $\omega$ is the eigenvalue of the  Sturm-Liouville problem resulting from the spatial and time symmetries of $\Psi$ as described in \cite{GuzmanUrena2004}.

The initial wave function for the collision of two configurations is the superposition of the wave functions of  two of these equilibrium configurations with different masses and linear momentum.
For the superposition we use the method in  \cite{AvilezGuzman2019}, specifically, we do not solve the  Sturm-Liouville problem for two equilibrium configurations with different masses. Instead, we exploit the scale invariance of the GPP system of equations \cite{GuzmanUrena2004}, namely that by scaling physical quantities as 
$t=\lambda^2 \hat{t}$, 
$x=\lambda \hat{x}$,
$\Psi = \hat{\Psi}/\lambda^2$, 
$V = \hat{V}/\lambda^2$, where $x$ represents any of the spatial coordinates and $\lambda$ is a number, the GPP system (\ref{eq:gpp}) remains unchanged. Thus, the solution of the GPP system for a given configuration means one can construct all other equilibrium configurations using this scaling property. In fact, a consequence of this scaling is that density and mass also scale as 
$\hat{\rho}=\lambda^4 \rho$, 
$\hat{M}=\lambda M$ which are important physical parameters of a scaled configuration used below.

In practice the typical equilibrium configuration is that with the central value of the wave function  $ \psi(r=0)=1$ that we will call $\psi(r)_1$ and has mass we call $M_1$. We choose one of the two configurations that will collide, to be precisely this standard configuration.

The second configuration that will collide, is a  configuration constructed with the scaling relations above, represented by the wave function $\psi(r)_{\lambda}=\lambda^2 \psi(r)_1$, with mass  $M_{\lambda} = \lambda M_{1}$. Notice that the scaling parameter happens to be the mass ratio  $\lambda = M_{\lambda}/ M_1 =\lambda = MR$ between the first and the second configurations used for the collision. In the analysis we consider the convention $0<\lambda<1$ in all cases, so that $M_{\lambda} < M_1$ always. 

\begin{figure}
\centering
\includegraphics[width= 7cm]{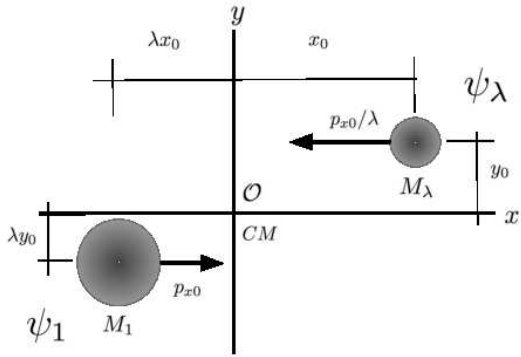}
\caption{Scheme of the initial conditions on the $xy-$plane, for the two configurations described by $\psi_1$ and $\psi_{\lambda}$. It illustrates the initial position and momentum in terms of the mass ratio $\lambda$, which are defined such that the center of mass is located at the origin and is expected to remain there.}
\label{fig:initialconditions}
\end{figure}

We then interpolate and superpose the two configurations in the numerical domain $D$. In order to maintain the system evolving within the numerical domain, we set the center of mass of the configuration at the coordinate origin. We parametrize the initial conditions by fixing the coordinates of the lighter configuration with mass $M_{\lambda}$ at $(x_0,y_0,0)$ with $x_0,y_0 >0$. Then, in order for the center of mass to lie at the origin, the center of the heavy configuration with mass $M_1$ must be centered at coordinates $(-\lambda x_0,-\lambda y_0,0)$. In this set up $y_0$ will play the role of impact parameter prior to merger.

The angular momentum is added through the imprint of linear momentum to the configurations along the $x$ direction only. For this we parametrize the momentum with the $x-$component of the heavy configuration with mass $M_1$ that we set to $p_{x0}$. Then again, in order to keep the center of mass approximately at the coordinate origin, the momentum of the light configuration must be $p_{x0}/\lambda$. The momentum is applied to each of the configurations by redefining $\psi_1\rightarrow e^{i p_{x0} x}\psi_1$ and $\psi_{\lambda} \rightarrow e^{-i  p_{x0}  x}\psi_{\lambda}$. Finally the wave function of the binary system at initial time is $\Psi = \psi_1 + \psi_{\lambda}$ and the scheme in Fig. \ref{fig:initialconditions} illustrates the initial conditions.

% ------------
\subsection{Diagnostics}

We monitor the dynamics of the system by evaluating some macroscopic variables. These include the mass $M$, kinetic energy $K$, gravitational energy $W$ and the $z$ component of the  angular momentum $L_z$. These quantities are 

\begin{eqnarray}
M &=& \int \Psi^{*}\Psi d^3 x \nonumber\\
K &=& -\frac{1}{2}\int \Psi^{\ast} \nabla^2 \Psi d^3x \nonumber\\
W &=& \frac{1}{2}\int \Psi^{\ast}V\Psi d^3x \nonumber\\
L_z &=& -i \int \Psi^{*} \left( x \frac{\partial \Psi}{\partial y} - y \frac{\partial \Psi}{\partial x}\right)  d^3 x
\label{eq:macroscopic}
\end{eqnarray}

\noindent where the integrals are calculated using the second order accurate trapezoidal rule. A first important quantity is the total energy $E=K+W$, whose sign determines when a system is bounded ($E<0$) or unbounded ($E>0$). A second one is $Q=2K+W$ which should be zero for a virialized system and allows one to determine when a system is near, tends to or is far from a virialized state. 

% -------------------------------------
% ----->     SECTION    <-----
% -------------------------------------
\section{Analysis}
\label{sec:analysis}

\subsection{Parameter space}

There is a garden variety of possible configurations that can be explored. However three parameters influence the behavior of the configuration resulting from the interaction between the two cores, namely,  mass ratio $MR$,  momentum $p_{x0}$ and parameter impact. These three parameters determine wide ranges of angular momentum and total energy values at initial time. It would be ideal to have the possibility of exploring a wide range of this parameter space. Nevertheless, due to the expensive computational cost of simulations, we restrict the exploration to illustrate the influence of some parameters using specific values.

First, we set two possible values of the mass ratio $MR=\lambda=0.5,1$, which are the equal mass scenario and the two to one mass ratio case, which will illustrate well the behavior of the system in unequal mass encounters.

Second, we consider various values of the impact parameter $y_0$. The radius of the configuration with mass $M_1$ is $r_{95} \sim 3.93$ in code units \cite{GuzmanUrena2004}, whereas that of mass $M_{1/2}$ is twice as big. Thus we study the range of values $y_0=1,2,...,10$ that accounts for scenarios ranging from nearly head-on to a separation various times bigger than the size of the structures.

Third, we distinguish between merger and unbounded scenarios.  In the first scenario the two configurations end up together and form a final configuration. In the second scenario, either the configurations flyby each other or behave as solitons. The momentum $p_{x0}$ is useful to generate the two scenarios, because it sets the amount on kinetic energy $K$ of the two configurations together. With a low value of this parameter the gravitational energy $W$ dominates, implying that $E<0$, otherwise a high momentum contributes to $K$ that can contribute importantly to the energy to be positive and produce unbounded configurations. The threshold value for the head-on scenario is found to be $p_{x0}\sim 0.7$ \cite{BernalGuzman2006} which serves as a guide to avoid non-merging cases.

We empirically found a range of values of $p_{x0}$ for which at initial time the total energy is negative for the two values of $MR$ and  all the values of $y_0$. Values in the range $p_{x0}\in[0,0.3]$ produce configurations  with negative energy. In what follows we use the case  $p_{x0}=0.1$ to illustrate the generic properties of mergers.

The values of these physical parameters suggest the numerical parameters to be used. The first parameter is the location  of the lighter configuration at $(x_0,y_0,0)$ with $x_0=10$ in all cases. We use this value because the interference at the origin between $\psi_1$ and $\psi_{\lambda}$, $\langle \psi_1,\psi_{\lambda}\rangle$ is less than $10^{-8}$. We consider the domain to be the box $D=[-40,40]^3$ and cover it with two refinement levels, and maximum resolution $\Delta x = 0.1r_{95}$ in the inner box, which covers the region where the dynamics is more important $D_h=[-20,20]^3$.

% ----- Subsection
\subsection{Global quantities}

In a merger scenario  the two cores collide and form a single final configuration whose density profile can eventually be fitted with a simple function that can be further used to understand and analyze the physics of different processes.

\begin{figure}
\includegraphics[width= 7cm]{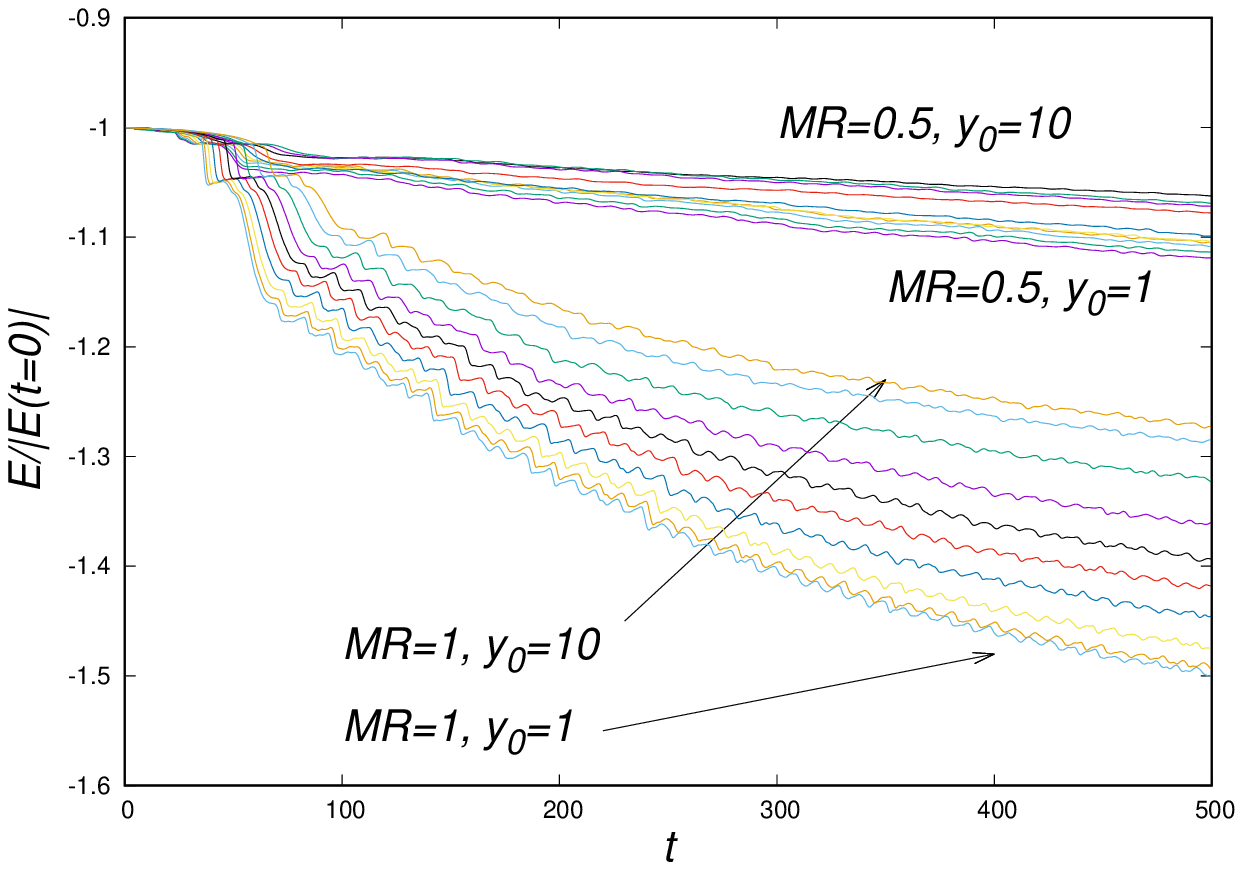}
\includegraphics[width= 7cm]{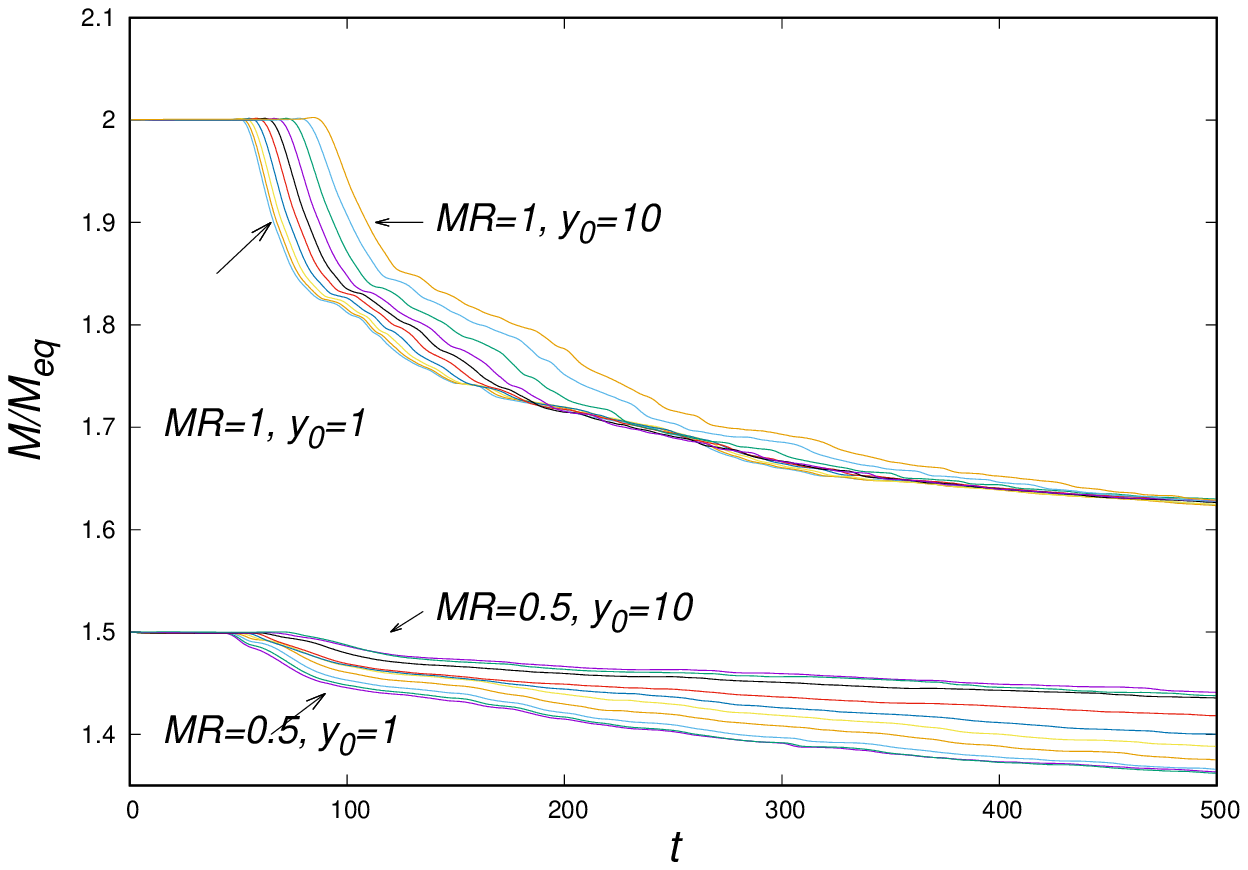}
\includegraphics[width= 7cm]{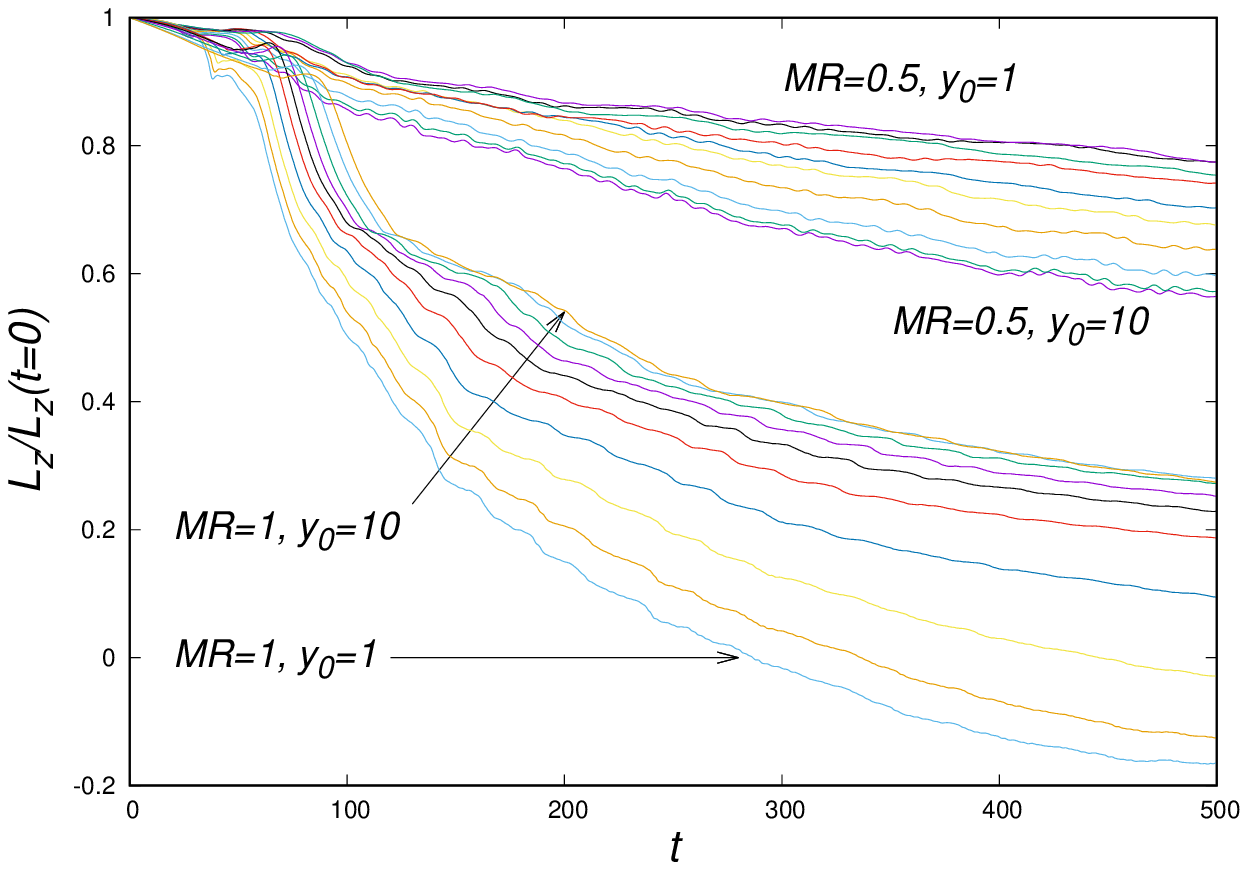}
\caption{For the  case $p_{x0}=0.1$, we show the total energy $E$, the total mass $M$ and $L_z$ for the two mass ratios considered $MR=0.5,~1$ and the ten impact parameter values $y_0=0,...,10$. Labels are used only for the two extreme values of $y_0=1,10$, whereas the unlabeled curves correspond to the other eight intermediate values of $y_0$.}
\label{fig:px0_1}
\end{figure}

We study now this scenario using $p_{x0}=0.1$ for the two values of $MR$ and all the values of the impact parameter $y_0$. The system of equations (\ref{eq:gpp}) is solved numerically for the initial conditions described above and we show the evolution of some of the scalars defined in Sec. II.C  in Fig. \ref{fig:px0_1}, for the ten values of $y_0=1,...,10$. 

The energy $E$ is shown normalized with the absolute value of its initial value. Notice that the total energy becomes more negative than at initial time, which indicates that the gravitational energy plays a more important role with time. The energy is also lost in a bigger proportion for smaller impact parameter $y_0$, and for $MR=1$ than for $MR=0.5$.

The mass is normalized with the mass of the standard equilibrium configuration $M_1$, therefore for $MR=1$ the total mass is initially $M=2$, whereas for $MR=0.5$ the initial mass is $M=1.5$. Notice that the mass decreases because matter is ejected and eventually captured by the numerical sponge. The combination of these two observations indicates that the mass lost during the process carries kinetic energy with it, exemplifying the Gravitational Cooling process \cite{SeidelSuen1991,GuzmanUrena2006}.

Notice also that for $MR=1$, the total mass is higher at initial time, but is also lost in a bigger percentage compared to the case of $MR=0.5$.  It can also be seen that the bigger the impact parameter $y_0$, the smaller the mass ejected during the process. For $MR=1$  the final mass converges to the same value independently of $y_0$, or equivalently to the initial angular momentum of the pre-merger configuration as discovered in \cite{Schwabe:2016}. Nevertheless for $MR=0.5$ this is not the case, at least within the time window of our simulations.

Another interesting result is that the matter also carries angular momentum with it. In the bottom panel of Fig. \ref{fig:px0_1}, the proportion of angular momentum during the merger is shown. 

The evolution of angular momentum shows an interesting behavior. For $MR=0.5$ the amount of $L_z$ released is between $\sim 20$\% for $y_0=1$ and $\sim 40$\% for $y_0=10$. In this sense, the simulations indicate that the merger process can produce final configurations with a wide range of values of angular momentum that could give origin to rotating galactic cores. However, for $MR=1$ the loss of angular momentum radiated away is of $\sim 65\%$ for $y_0=10$ and even turns negative for $y_0=1,2,3$ in the time window of the simulations and could hold also for other values in a bigger time domain. This result is interesting and the reason for the change of sign is that small values of $y_0$ correspond to nearly head-on situations. Since $y_0$ is the value of the impact parameter only at the center of each configuration, part of the matter ejected should be that initially located farther from the $x-$axis which carries angular momentum with it when it abandons the domain. This turn in the direction of rotation could be an interesting sign that eventually may provide restrictions to the model or predictions.

% ----- Subsection
\subsection{Equal mass case}

\begin{figure}
\centering
\includegraphics[width= 4.25cm]{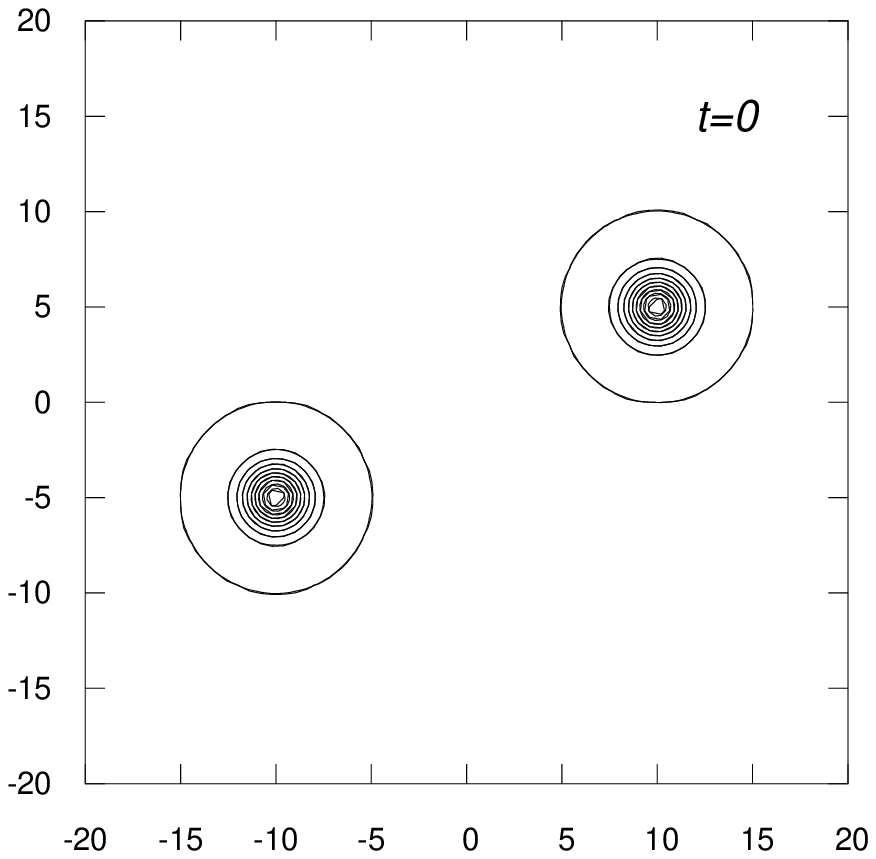}
\includegraphics[width= 4.25cm]{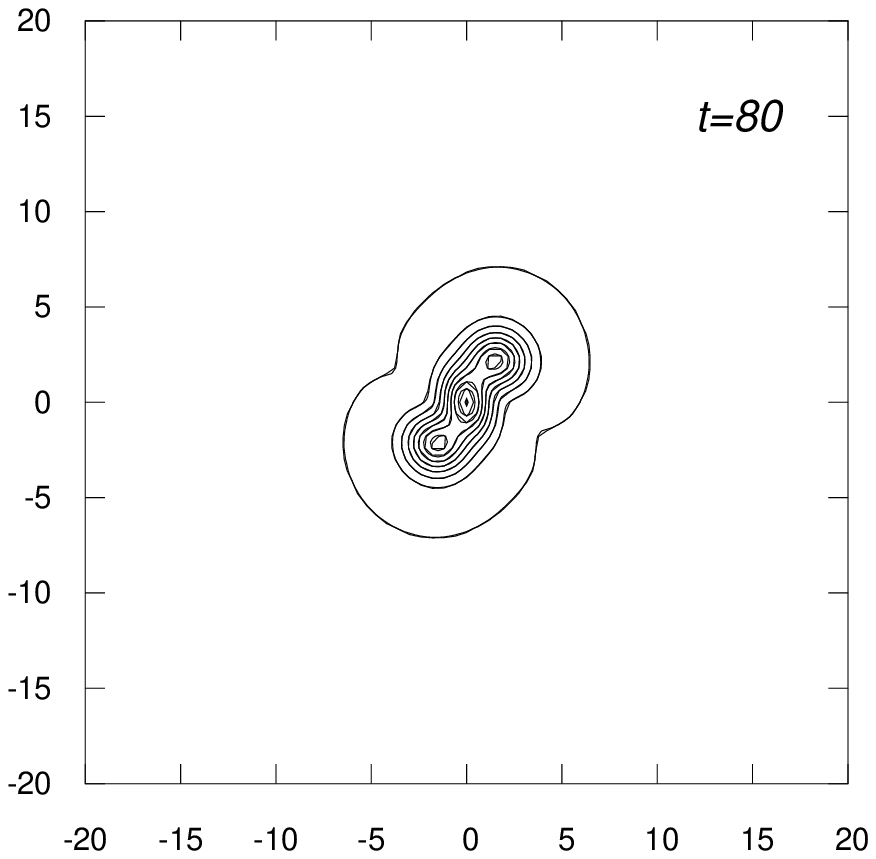}
\includegraphics[width= 4.25cm]{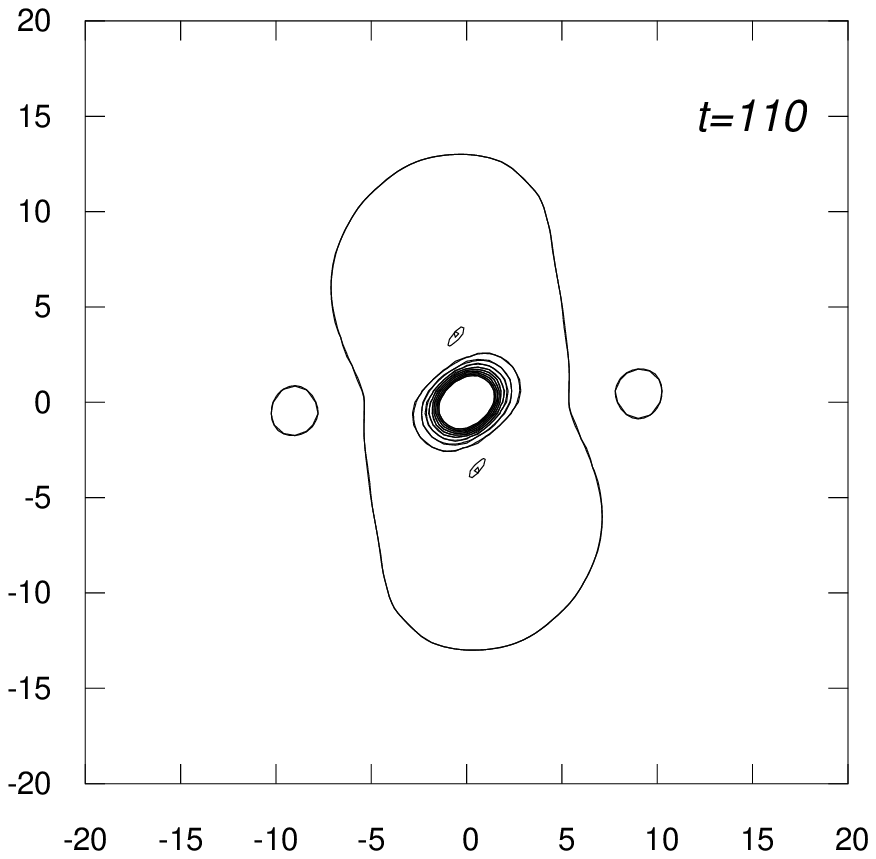}
\includegraphics[width= 4.25cm]{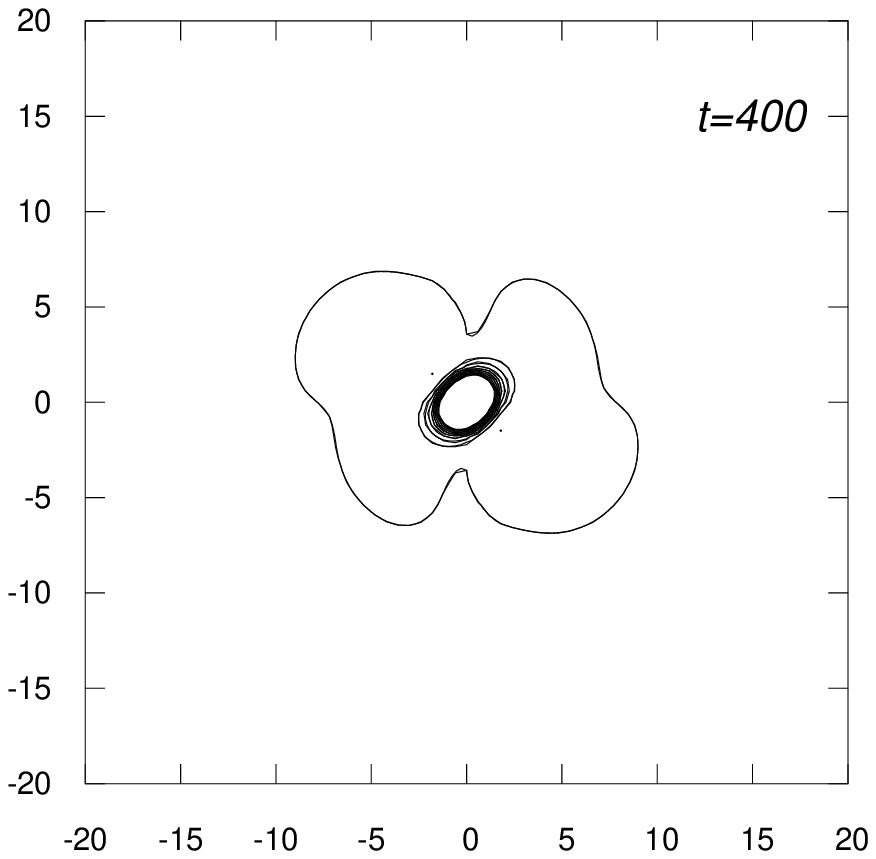}
\includegraphics[width= 4.25cm]{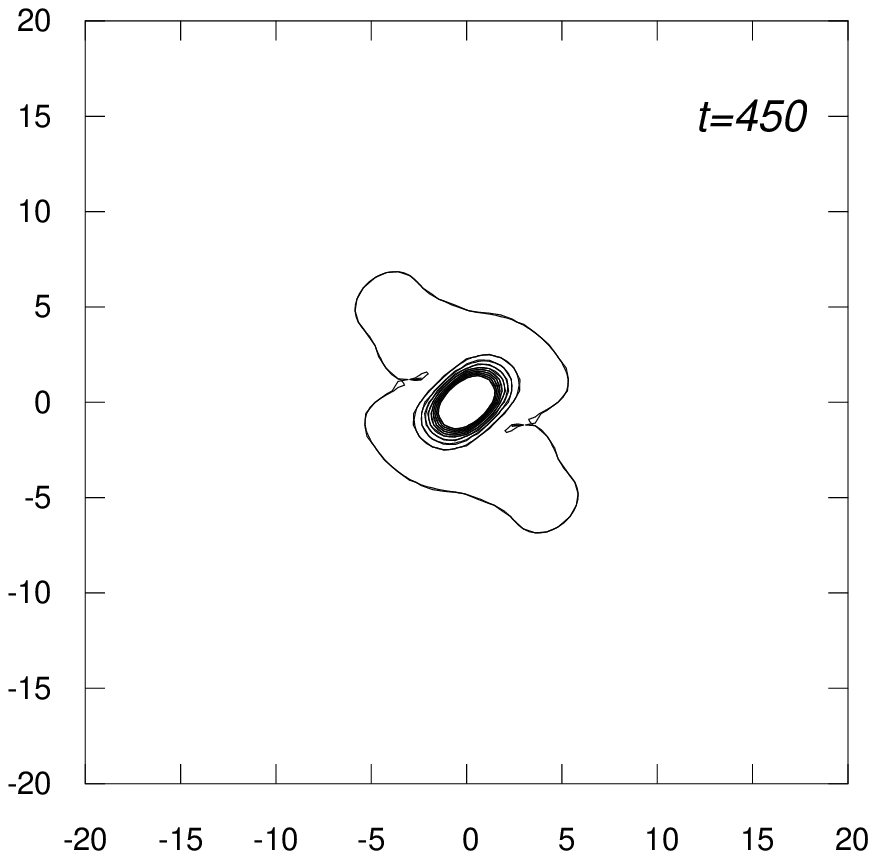}
\includegraphics[width= 4.25cm]{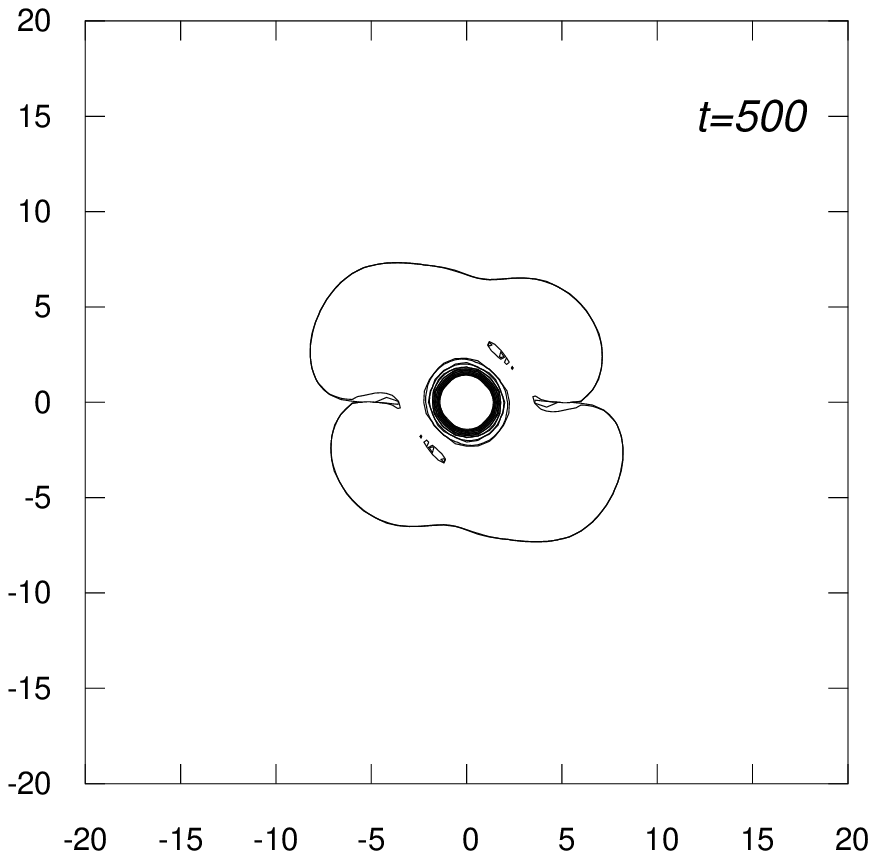}
\caption{Density contours on the $xy-$plane for the equal mass merger with $p_{x0}=0.1$ and $y_0=5$.}
\label{fig:isocurves}
\end{figure}

The evolution of a specific simulation is shown in Fig. \ref{fig:isocurves} for the equal mass case. The final configuration remains centered at the coordinate origin, rotates and has an ellipsoidal density profile. Animations of this and cases with various other parameter values are available in the supplemental material \cite{suppl}.

\begin{figure}
\includegraphics[width= 7cm]{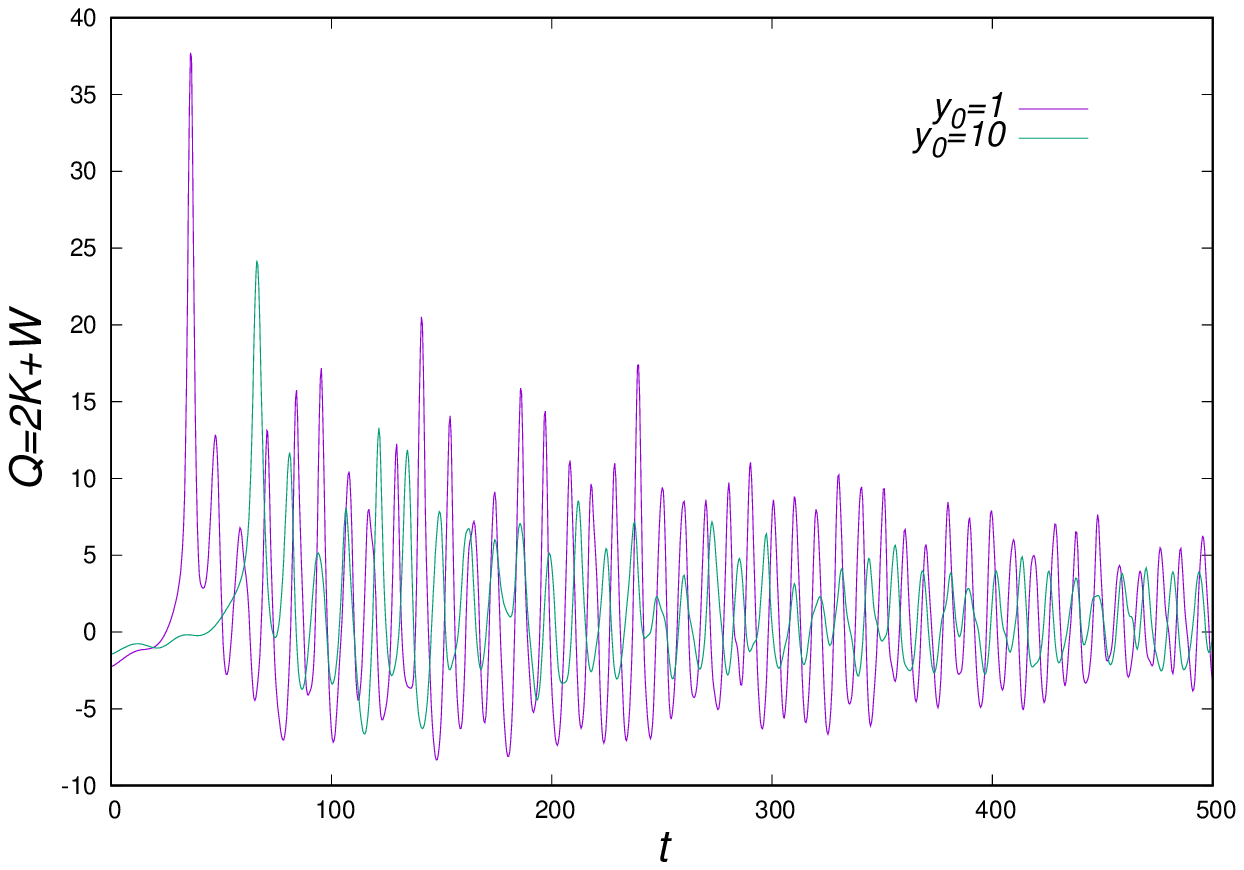}
\includegraphics[width= 7cm]{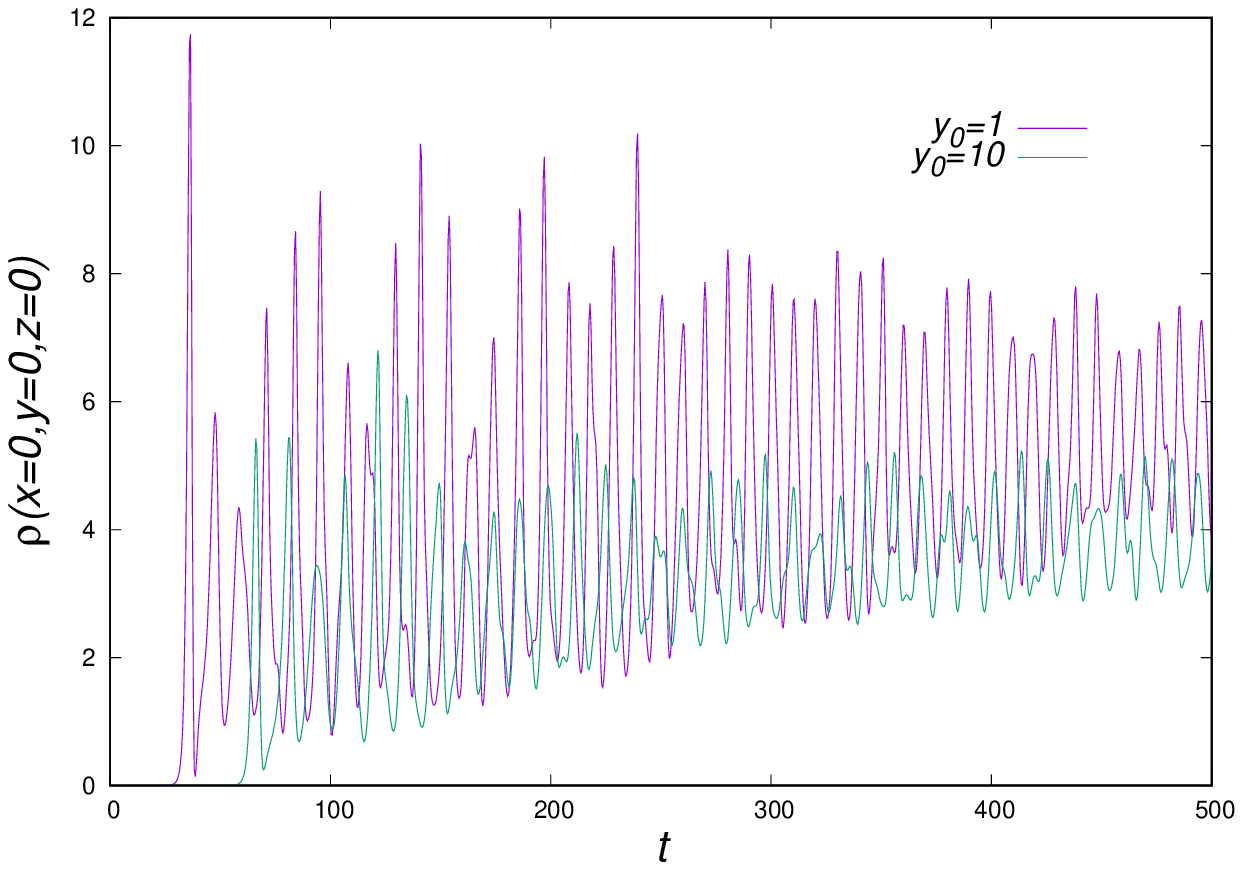}
\caption{For the  case $p_{x0}=0.1$ and $MR=1$, we show $Q=2K+W$ and the central value of the density as a function of time for $y_0=0$ and $y_0=10$. Oscillations are smaller for $y_0=10$ than for the nearly head-on case $y_0$=1.}
\label{fig:relaxationpx0_1}
\end{figure}

In order to learn more about the dynamical behavior of the final configuration, we track the value of the central density and $Q=2K+W$ as functions of time that are shown in Fig. \ref{fig:relaxationpx0_1} for the two extreme values of the impact parameter $y_0=1,~10$. It can be seen that the quantity $Q$ oscillates around zero with a decreasing amplitude as expected for the Gravitational Cooling \cite{AvilezGuzman2019}. 

\begin{figure}
\includegraphics[width= 7cm]{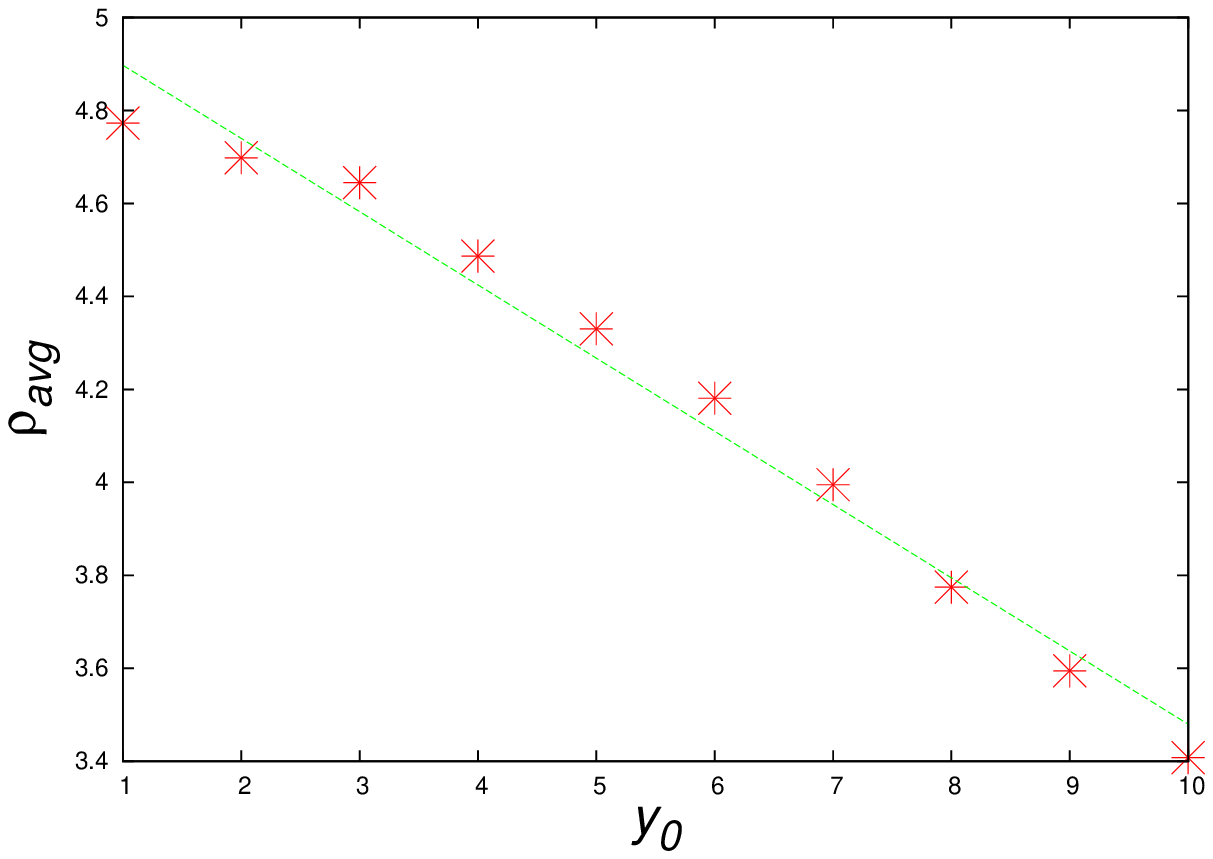}
\includegraphics[width= 7cm]{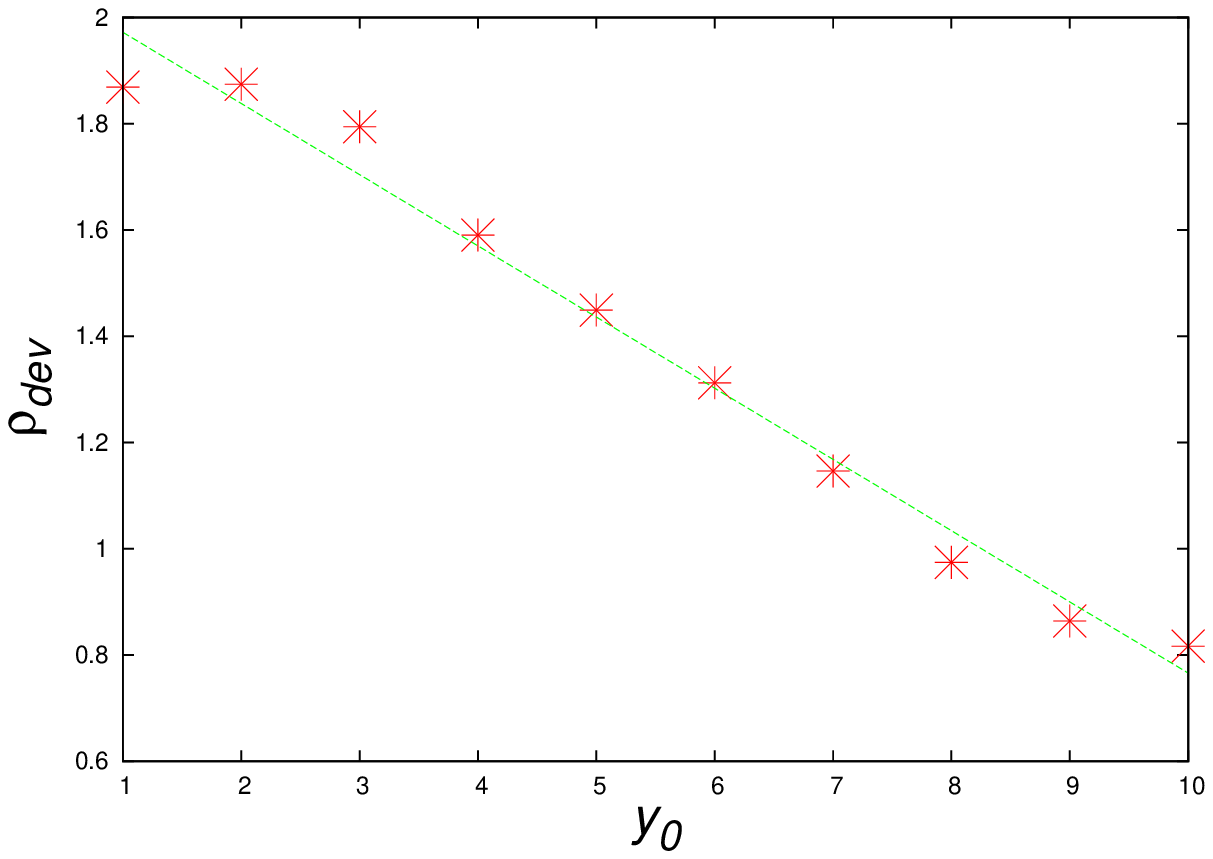}
\includegraphics[width= 7cm]{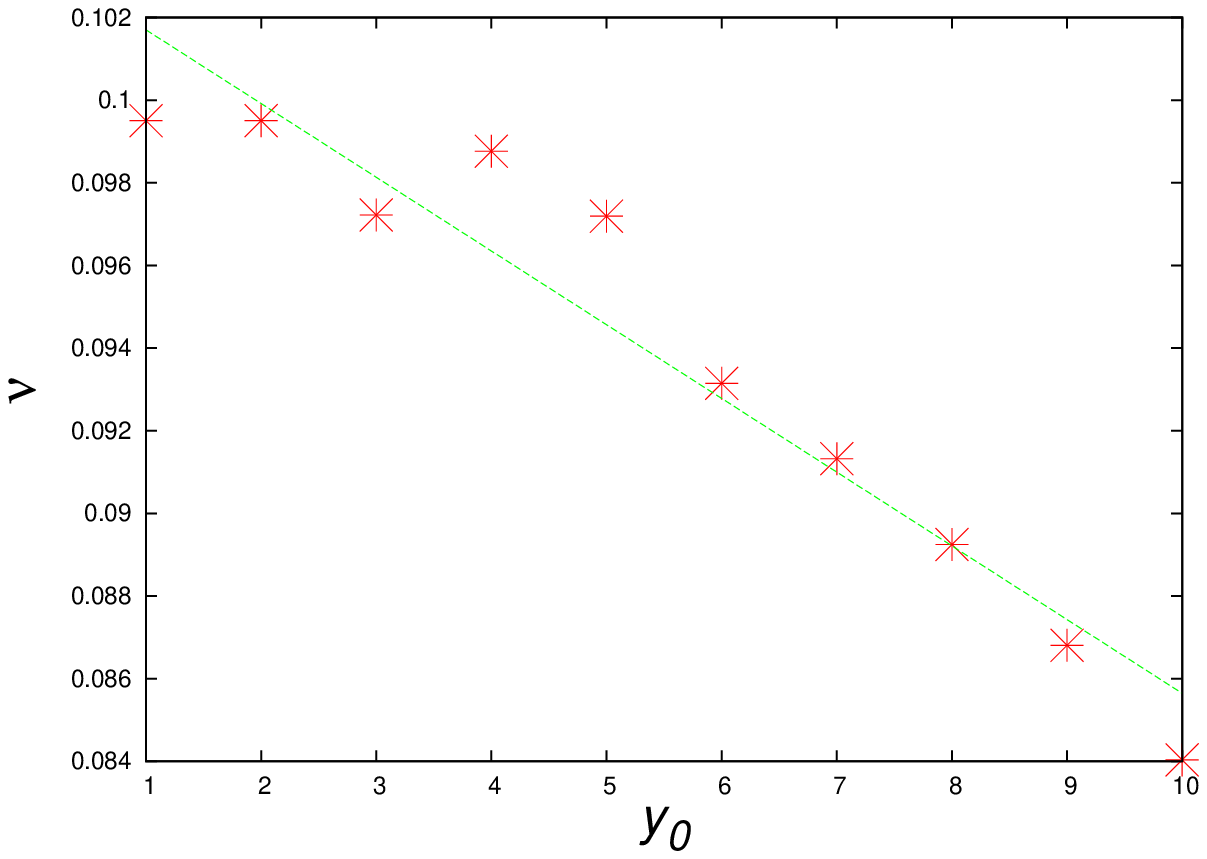}
\caption{For the  case $p_{x0}=0.1$, $MR=1$, the stars indicate the average in time of the central density of the final configuration $\rho_{avg}$, its standard deviation $\rho_{dev}$ and the peak frequency $\nu$ for the ten values of $y_0$ used. For each of these quantities we show a linear fit, suggesting the dependency on $y_0$ can be linear.}
\label{fig:statistics}
\end{figure}

The central density oscillates changing values by factors between two and three in the nearly head-on case $y_0=1$ and smaller oscillations for $y_0=10$. Fig. \ref{fig:relaxationpx0_1} suggests that the amplitude of the oscillations and the central value of the density depend on the impact parameter. In order to find a dependency on $y_0$ we calculated the average density $\rho_{avg}$ and its standard deviation to have a measure of the amplitude variation around the average $\rho_{dev}$, for $t>200$. The results are shown in Fig. \ref{fig:statistics}, which suggest that both, the central density and oscillation amplitude depend on $y_0$ linearly. Finally, calculating a Fourier Transform within the same time domain, we obtain the peak frequency associated to the dominant density oscillation mode, which also depends on the impact parameter as shown in the third panel of Fig. \ref{fig:statistics}. Knowing that the final mass is the same for all values of $y_0$, the oscillation frequency is genuinely different for different values of $y_0$.

In structure formation simulations \cite{Schive:2014,Schive:2014hza,Schwabe:2016,Niemeyer2016,Mocz2017} the density distributions resulting from the interaction of two or more configurations are associated to density profiles with a solitonic core and a tail, however it is  not quite specified whether these are final, relaxed configurations or not. As far as we can tell, the oscillations shown in Fig. \ref{fig:relaxationpx0_1} do not correspond to a relaxed structure. Even though the density profile can be fitted with the core profile

\begin{equation}
\rho_{soliton}(r) = \rho_0 \left[ 1 + 0.091 \left( \frac{r}{r_c}\right)^2\right]^{-8}
\label{eq:solitonic}
\end{equation}

\noindent as indicated in \cite{Schive:2014,Mocz2017}, where $\rho_0$ is the central density and $r_c$ is a core radius. 

\begin{figure}
\includegraphics[width= 7cm]{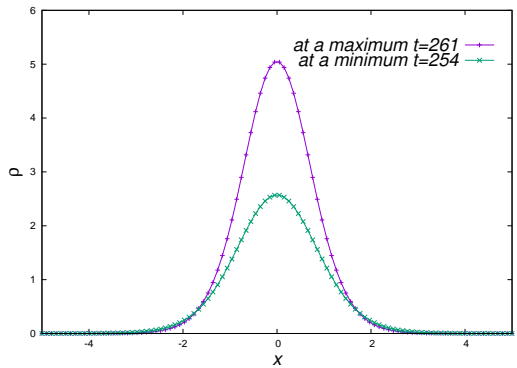}
\caption{Density profile at two different times for the  case $p_{x0}=0.1$, $MR=1$ and $y_0=10$. The dynamics can be seen in the corresponding animation within supplemental material \cite{suppl}.}
\label{fig:densityprofile}
\end{figure}

The issue is that the density is oscillating with considerable amplitude as seen in Fig. \ref{fig:relaxationpx0_1}. Nevertheless, the fitting was performed on the density profile when the central density is at a local maximum and at a local consecutive minimum. The fitting parameters results appear in Table \ref{tab:fits} for the projection of $\rho$ along the $x-$axis. Notice that the central density changes by a factor between two and three from a minimum to a maximum, whereas the core radius  changes by nearly 50\%. An example of how the density profile changes in time is illustrated in Fig. \ref{fig:densityprofile}, where the projection of the density along $x$ is shown at two specific times for $y_0=10$ at a minimum ($t=254$) and at a maximum ($t=261$).

\begin{table}
\begin{tabular}{|c|c|c|c|c|}\hline
$y_0$ 	&	$\rho_0$ 	& $r_c$ 	& $t$ 	& comment			 \\\hline\hline
1		&	8.280	& 0.647	& 351	& density at a maximum 		\\
		&	2.684	& 0.978	& 358	& density at a minimum 	\\
5		&	6.862	& 0.776	& 348	& density at a maximum			\\
		&	2.566	& 0.982	& 355	& density at a minimum	\\
10		&	5.053	& 0.852	& 254	& density at a maximum				\\
		&	2.571	& 4.023	& 261	& density at a minimum	\\\hline
\end{tabular}
\caption{Fitting parameters of $\rho(x,0,0)$ for the case $MR=1$, $p_{x0}=0.1$ and three values of the impact parameter $y_0$.}
\label{tab:fits}
\end{table}

In order to have an idea of the physical time scale of these oscillations, we use the recipe in \cite{Mocz2017}. Considering a boson mass value $2.5\times 10^{-22}$eV and that core radius of the final configuration is converging to $r_c=1$kpc, using the range of frequencies $\nu\in(0.084,0.1)$ from Fig. \ref{fig:statistics},  the period of the density oscillations is in the range $T \sim 0.76 - 0.91$Gyr. If the core radius is considered to be $r_c=0.25$kpc the period is within the range $T\sim 47 - 57 $Myr.

% ----- Subsection
\subsection{Unequal mass case}

As described before, the case we look at in detail corresponds to $MR=0.5$. The time dependence of $M$, $Q$ and $L_z$ appears in Fig. \ref{fig:px0_1}. General properties are very similar to those of the equal mass case. The loss of mass and angular momentum is smaller when the impact parameter is bigger.

\begin{figure}
\centering
\includegraphics[width= 4.25cm]{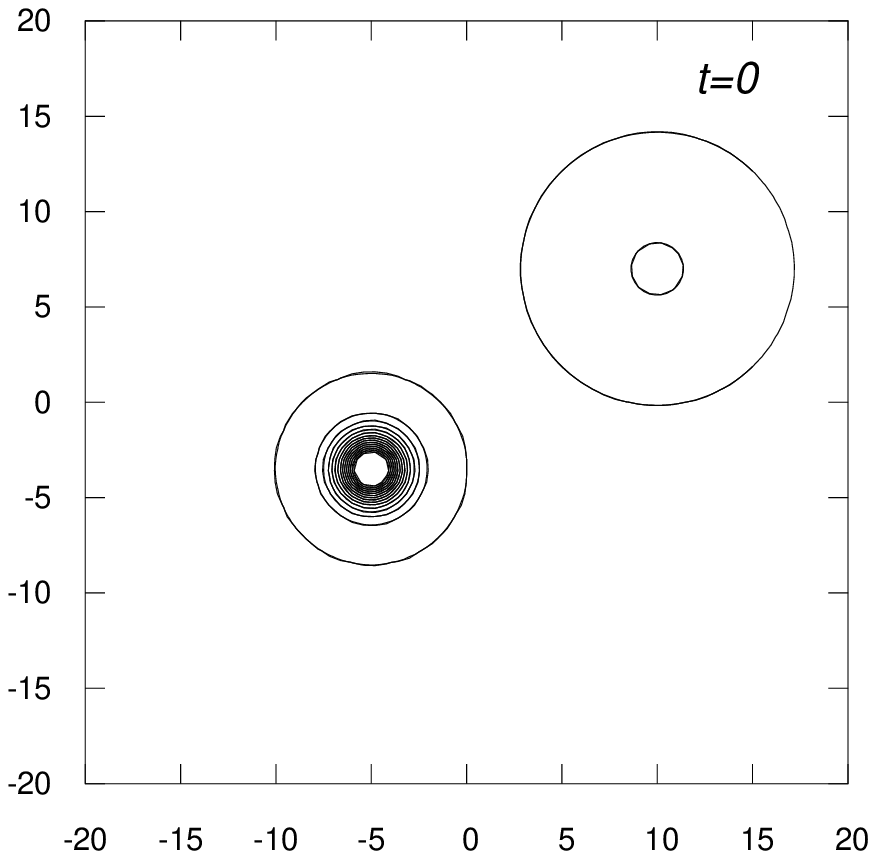}
\includegraphics[width= 4.25cm]{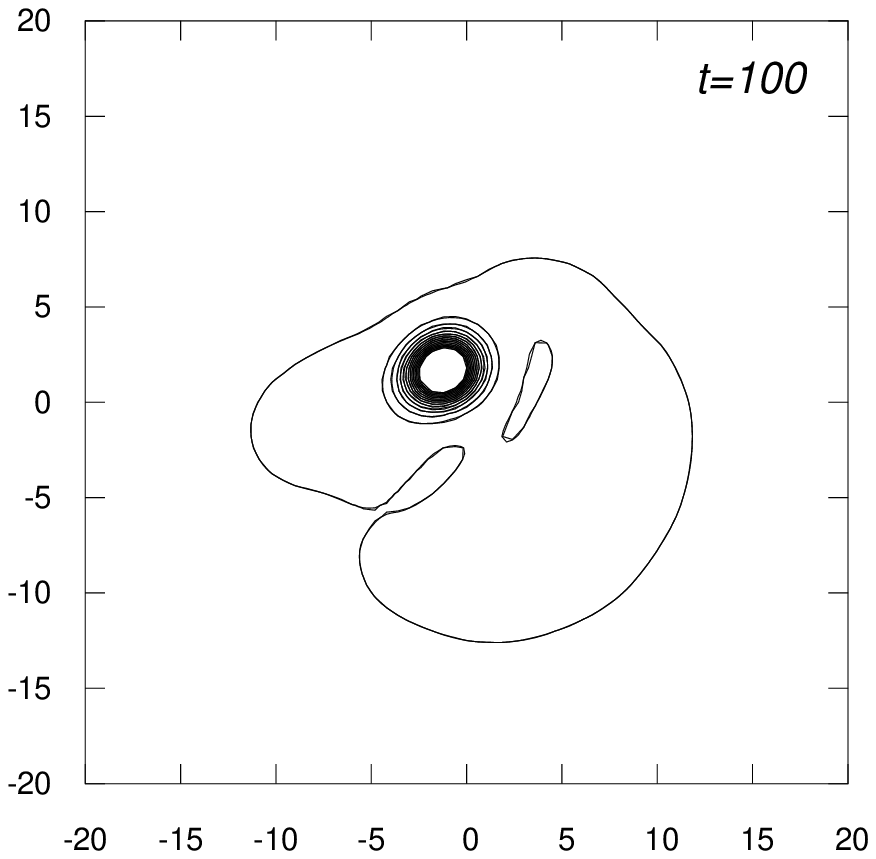}
\includegraphics[width= 4.25cm]{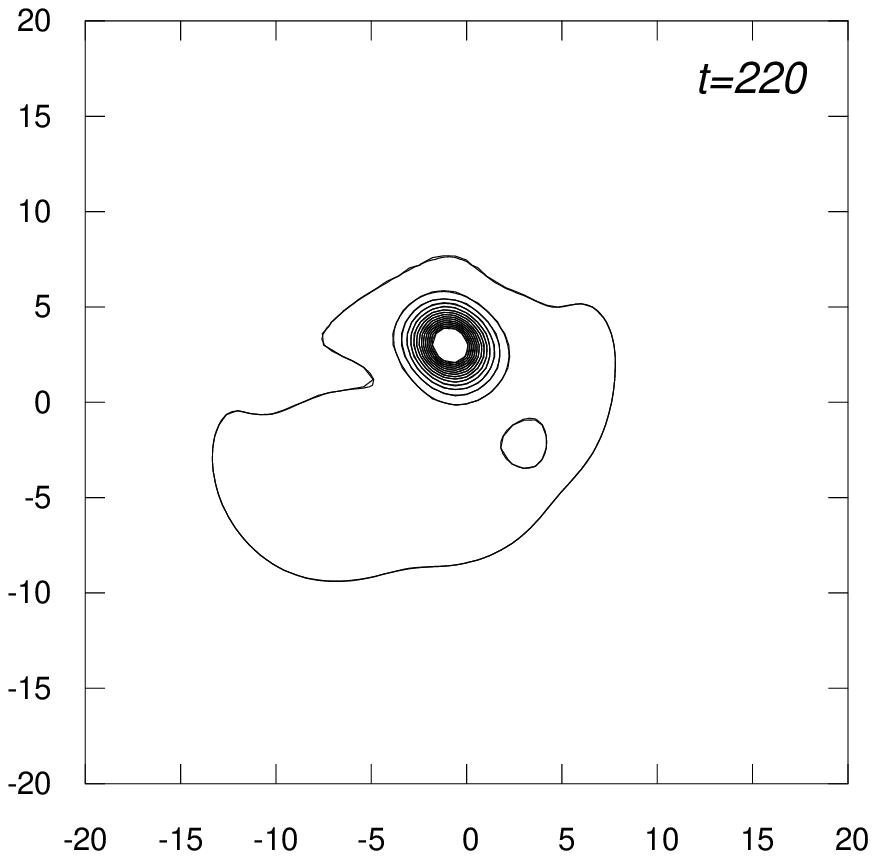}
\includegraphics[width= 4.25cm]{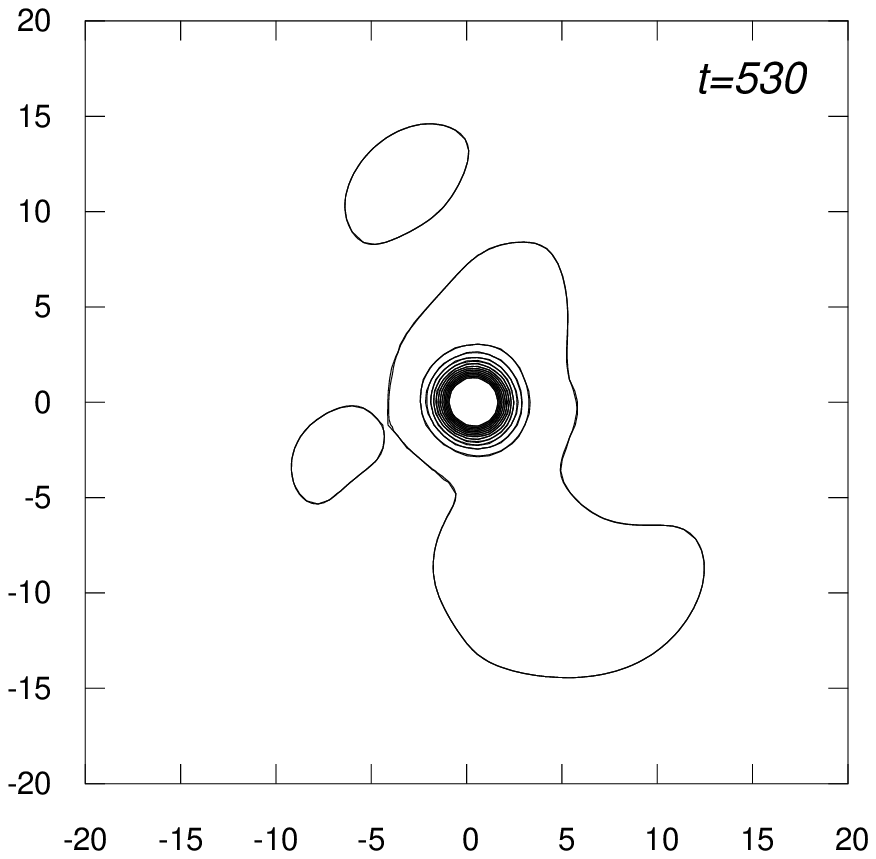}
\includegraphics[width= 4.25cm]{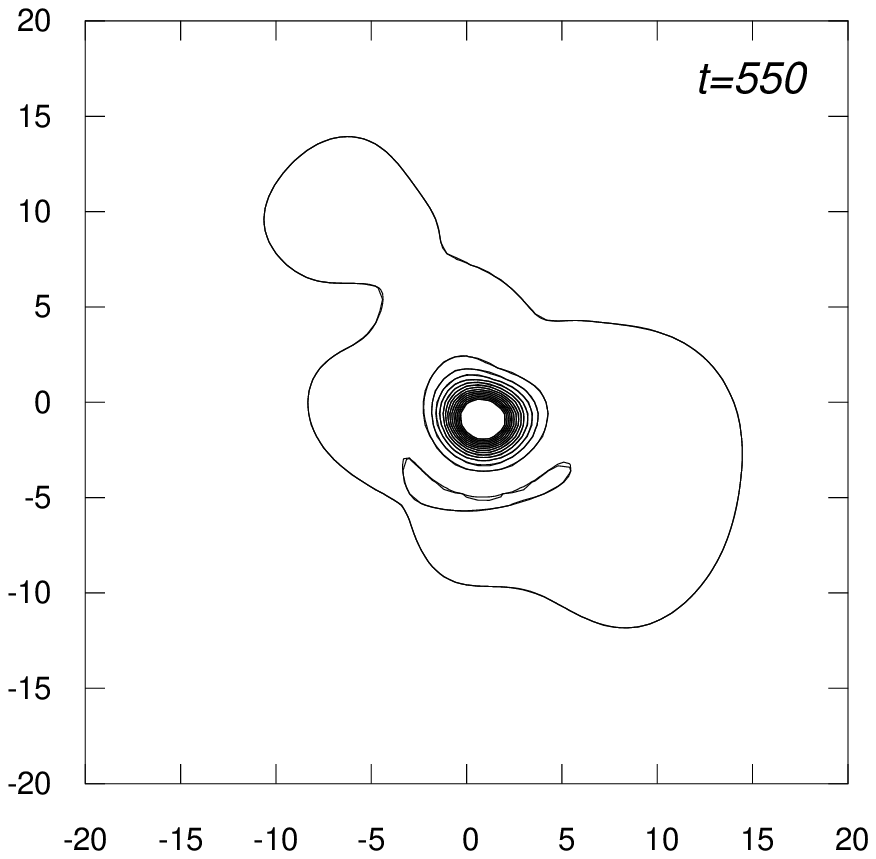}
\includegraphics[width= 4.25cm]{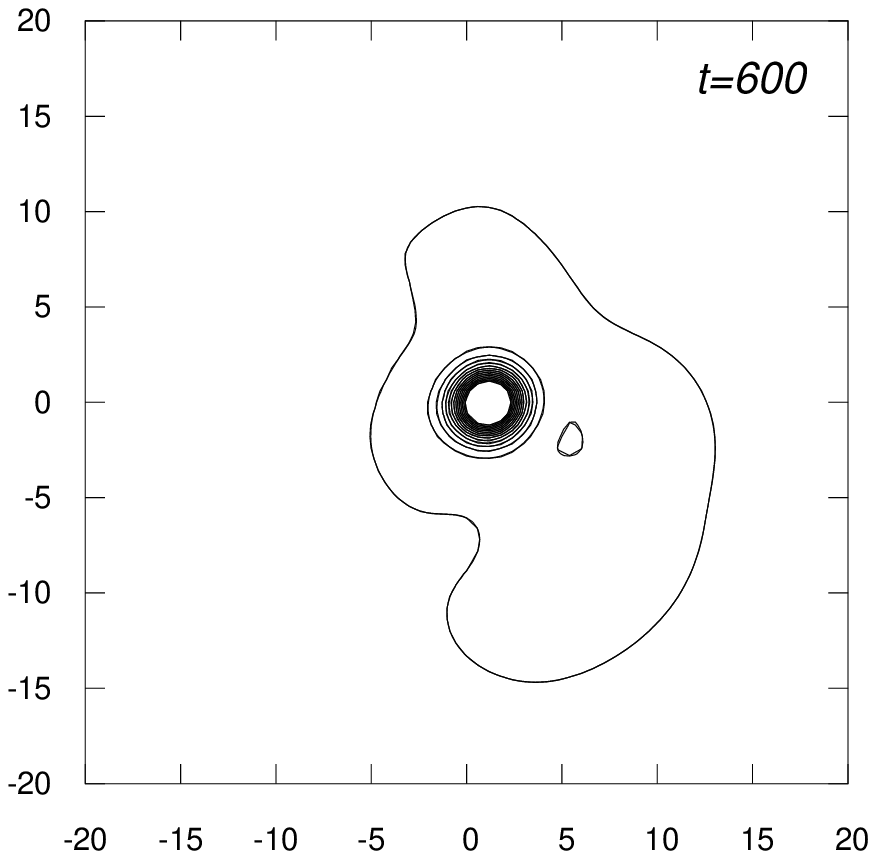}
\caption{Density isocontours on the $xy-$plane for the collision for $MR=0.5$ with $p_{x0}=0.1$ and $y_0=7$.}
\label{fig:isocurves2}
\end{figure}

Snapshots of the unequal mass $MR=0.5$ merger with $y_0=7$ are presented in Fig. \ref{fig:isocurves2}. The resulting high density region wobbles around the origin due to the asymmetric distribution of matter and at some point evolves toward the coordinate origin. Animations for other values of the parameters are also shown in the supplemental material \cite{suppl}.

\begin{figure}
\includegraphics[width= 7cm]{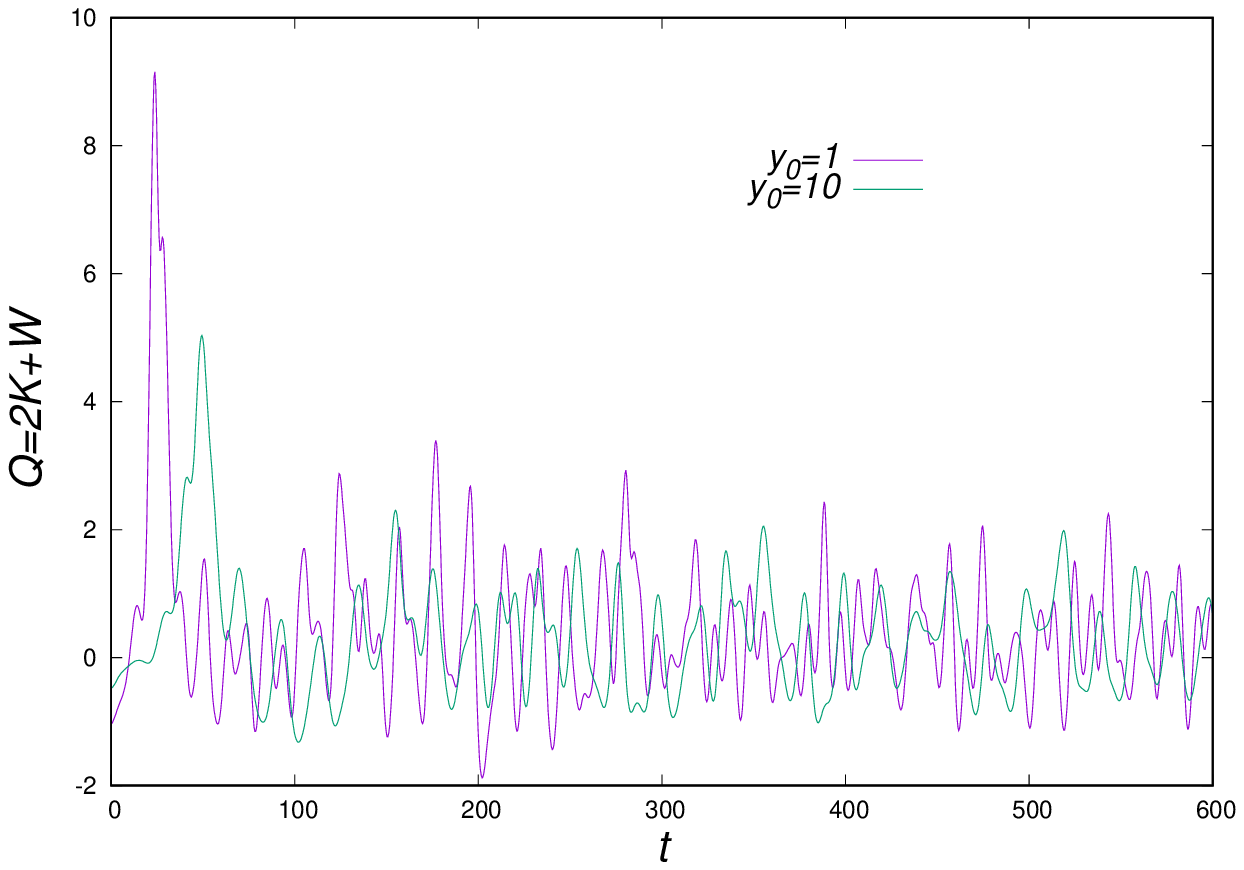}
\includegraphics[width= 7cm]{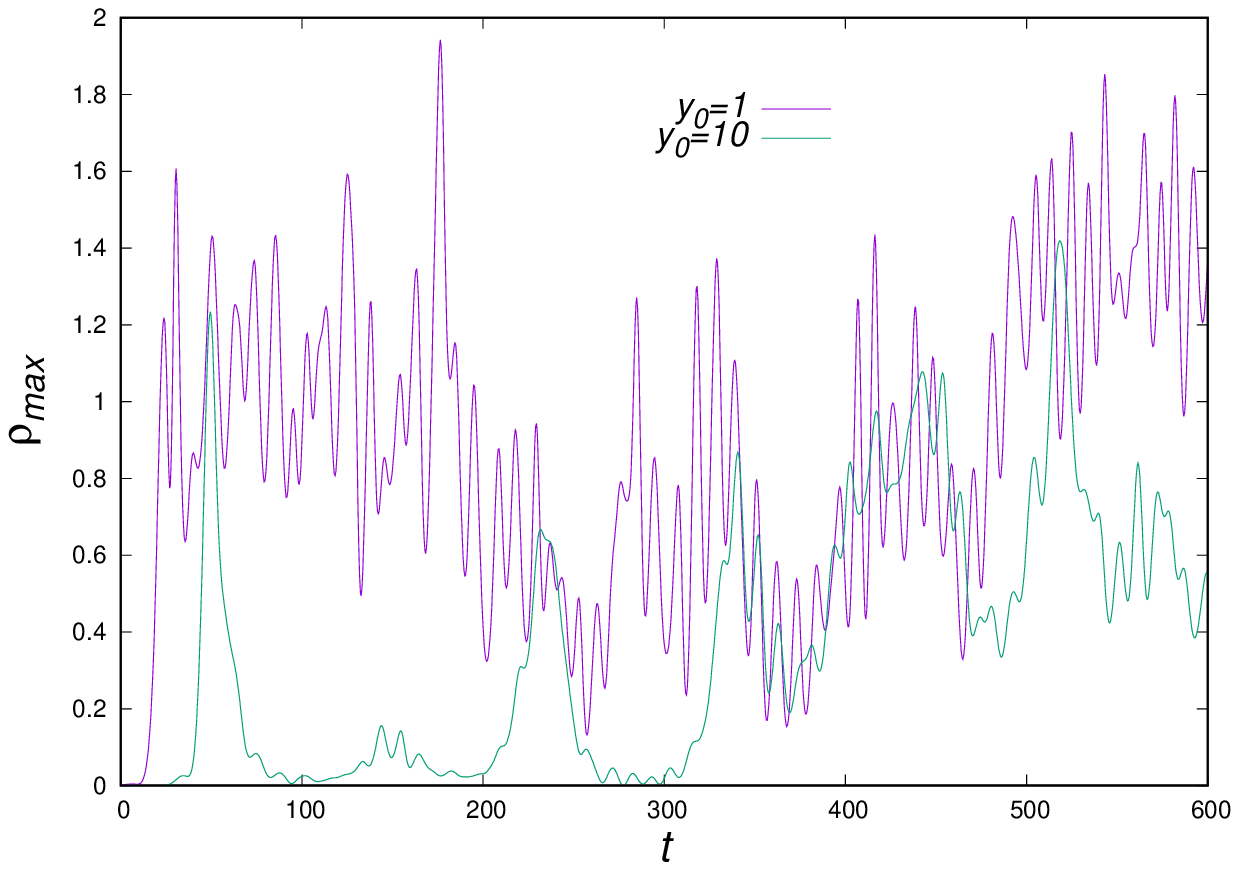}
\caption{For the  case $p_{x0}=0.1$ and $MR=0.5$, we show $Q=2K+W$ and the maximum value of the density as a function of time for the cases $y_0=1$ and $y_0=10$.}
\label{fig:relaxationpx0_1MR05}
\end{figure}

What is different from the equal mass case is the relaxation process. The evolution of $Q=2K+W$ and the central value of the density are shown in Fig. \ref{fig:relaxationpx0_1MR05} for the two extreme values of the impact parameter $y_0=1,10$. The value of $Q$ oscillates around zero with amplitude an order of magnitude smaller than in the equal mass case. For density on the other hand, since the configuration is wobbling around the coordinate origin, instead of tracking the central value of the density we track its maximum value $\rho_{max}$. The result in the Figure is generic behavior for the unequal mass cases with values of $MR$ between 0.5 and 1 we experimented with. The highly dynamical behavior is due to the fact that the small configuration with mass $M_{\lambda}$ approaches with a higher velocity and the distribution is much less symmetric than for $MR=1$. This explains a quick ejection of kinetic energy so that $Q$ acquires small values.

The density does not show any clear sign of relaxation or a particular dominant mode during the time window used in the simulations. This is perhaps a major obstacle when the density is fitted with a space-dependent density fitting function. Unlike the equal mass case, where the average of the density is a good estimate of the asymptotic value, here the expected value of the central density is uncertain. Nevertheless, Fig. \ref{fig:relaxationpx0_1MR05} indicates that the central density of the final configuration depends on the impact parameter $y_0$.

% -------------------------------------
% ----->     SECTION    <-----
% -------------------------------------
\section{Conclusions and discussion}
\label{sec:conclusions}

We have presented the merger process of ultralight bosonic dark matter cores, with detailed illustrations of the equal mass case $MR=1$ and a representative unequal mass case $MR=0.5$.

In the equal mass case it was found that the final configuration oscillates with amplitudes that depend on the parameters of the binary prior to merger, namely, the mass ratio of the two initial cores, linear momentum and impact parameter. The resulting final configuration was fitted with a solitonic density profile at different times during the relaxation process. It was found that the density may change by factors of nearly three whereas the core radius can change by nearly 50\% percent, and that the amplitude and frequency of the oscillations can be linearly related to the impact parameter of the merger.

In the unequal mass case, due to the size of the initial configurations, the interference becomes important in the symmetry of the final high density zone, which wobbles around the center of mass before it settles toward a nearly fixed location. The density in this case oscillates, however with an irregular superposition of modes, although with values of $Q$ indicating that globally the system evolves around a virialized state. 

In both scenarios, it calls the attention the fact that the density seems far from a stationary state in cosmological time scales. The reason is that the amplitude of oscillations of the configuration resulting from a merger is not small, and perhaps it would be useful to consider time averages in such fittings.

In order to determine observational restrictions of this dark matter model, it seems unavoidable to systematically analyze the effect of the dynamics of a configuration resulting from a merger on the luminous matter that can be involved in the process. For example their survival questioned for specific scenarios of the head-on case in \cite{GonzalezGuzman2016} or restrictions from the existence of star clusters near galactic cores \cite{Marsh2018}.

% ----->     ACKNOWLEDGMENTS  

\section*{Acknowledgments}
This research is supported by Grants CIC-UMSNH-4.9, CIC-UMSNH-4.23, CONACyT 258726 
(Fondo Sectorial de Investigaci\'on para la Educaci\'on). The runs  were carried out in the computer farm funded by CONACyT 106466 and the Big Mamma cluster at the IFM.

% ----->     REFERENCES     


\begin{thebibliography}{99}

\bibitem{Matos-Urena:2000} T. Matos and L. A. Ure\~na-L\'opez, Classical Quantum Gravity 17, L75 (2000)

\bibitem{Ostriker:2016} L. Hui, J. P. Ostriker, S. Tremaine, and E. Witten, Phys. Rev. D 95, 043541 (2017)

\bibitem{Chen:2016} S-R. Chen, H-Y. Schive, T. Chiueh, MNRAS 468, 1338 - 1348 (2017)

\bibitem{Schive:2015} H-Y. Schive, T. Chiueh, Tzihong, T. Broadhurst, K-W. Huang, ApJ 818, 89 (2016)

\bibitem{Du:2016} X. Du, C. Behrens, J. C. Niemeyer, B. Schwabe, Phys. Rev. D 95, 043519 (2017)

\bibitem{Velmaat2018} J. Velmaat, J. C. Niemeyer, B. Schwabe, Phys. Rev. D 98, 043509 (2018)

\bibitem{Schive:2014} H-Y. Schive, T. Chiueh, T.  Broadhurst, Nature Phys. 10, 496-499 (2014)

\bibitem{Schive:2014hza} H-Y. Schive, M-H. Liao, T-P. Woo, S-W. Wong, T. Chiueh, T. Broadhurst, W-Y. P. Huang, Phys. Rev. Lett. 113, 261302 (2014)

\bibitem{Marsh-Ferreira:2010} D. J. Marsh, P. G. Ferreira, Phys. Rev. D 82, 103528 (2010)

\bibitem{GuzmanUrena2003} F. S. Guzm\'an and L. A. Ure\~na-L\'opez, Phys. Rev. D 68, 024023 (2003)

\bibitem{GuzmanUrena2006} F. S. Guzm\'an and L. A. Ure\~na-L\'opez, Astrophys. J. 645, 814 (2006).

\bibitem{Mocz2017} P. Mocz, M. Vogelsberger, V. Robles, J. Zavala, M. Boylan-Kolchin, A. Fialkov, L. Hernquist 
MNRAS 471, 4559 - 4570 (2017)

\bibitem{ChavanisDamping} P-H. Chavanis, Eur. Phys. J. Plus, 132, 248 (2017)

\bibitem{AvilezGuzman2019} A. A. Avilez and F. S. Guzm\'an, Phys. Rev. D 99, 043542 (2019)

\bibitem{GonzalezGuzman2016} F. S. Guzm\'an, J. A. Gonz\'alez, and J. P. Cruz-P\'erez, Phys. Rev. D 93, 103535 (2016)

\bibitem{Marsh2018} D. J. E. Marsh, J. C. Niemeyer, e-print: arXiv:1810.08543

\bibitem{BernalGuzman2006b} A. Bernal and F. S. Guzm\'an, Phys. Rev.D 74, 063504 (2006)

\bibitem{Niemeyer2016} J. Velmaat and J. Niemeyer, Phys. Rev. D 94, 123523 (2016)

\bibitem{Schwabe:2016}  B. Schwabe, J. C. Niemeyer, J. F. Engels, Phys. Rev. D 94, 043513 (2016)

\bibitem{ChavanisCore} P-H. Chavanis, e-print: arXiv: 1810.08948

\bibitem{BEC2014}F. S. Guzm\'an, F. D. Lora-Clavijo, J. J. Gonz\'alez-Avil\'es, F. J. Rivera-Paleo. Phys. Rev. D 89, 063507 (2014)

\bibitem{BEC2015} F. S. Guzm\'an, F. D. Lora-Clavijo, Gen. Rel. Grav. 47, 21 (2015)

\bibitem{Tula} T. Bernal, L. M. Fern\'andez-Hern\'andez, T. Matos, M. A. Rodr\'iguez-Meza,  MNRAS 475, 1447-1468 (2016)

\bibitem{RindlerShapiro2012} T. Rindler-Daller and P. R. Shapiro, Mon. Not. R. Astron. Soc. 422, 135 (2012)

\bibitem{DavidsonSchwetz2016} S. Davidson, T. Schwetz, Phys. Rev. D 93, 123509 (2016)


\bibitem{Hertzberg2018} M. P. Hertzberg, E. D. Schiappacasse, JCAP 08, 028 (2018)

\bibitem{ParedesMichinel2016} A. Paredes, H. Michinel. Physics of the Dark Universe, volume 12 (2016), pages 50-55.

\bibitem{BergerOliger} M. J. Berger and J. Oliger, J. Comp. Phys. 53, 484 (1984)

\bibitem{GuzmanUrena2004} F. S. Guzm\'an and L. A. Ure\~na-L\'opez, Phys. Rev. D 69, 124033 (2004)

\bibitem{BernalGuzman2006} A. Bernal and F. S. Guzm\'an, Phys. Rev. D 74, 103002 (2006)

\bibitem{SeidelSuen1991} E. Seidel, W-M Suen, W-M. Phys. Rev. Lett., 66, 1659 (1991)

\bibitem{suppl} http://www.ifm.umich.mx/\~{}guzman/Supplemental/
%ÑÑÑ. 1994, Phys. Rev. Lett., 72, 2516

%\bibitem{Guzman2019} F. S. Guzm\'an, Phys. Rev. D 99, 083513 (2019)


\end{thebibliography}
\end{document}